\documentclass[11pt]{article}
\begin{document}
~\hfill{\footnotesize UICHEP-TH/99-2 , IP/BBSR/99-5~~\today}
\vspace{.3in}
\begin{center}
{\Large {\Large \bf Comment on ``Self-Isospectral Periodic Potentials and Supersymmetric Quantum Mechanics"}}
\end{center}
\vspace{.3in}
\begin{center}
{Uday Sukhatme}\\
{Department of Physics, University of Illinois at Chicago, Chicago, Illinois 60607}\\
\vspace{.2in}
{Avinash Khare}\\
{Institute of Physics, Sachivalaya Marg, Bhubaneswar 751005, Orissa, India}\\
\end{center}
\vspace{.3in}
\begin{abstract}
We show that the formalism of supersymmetric quantum mechanics applied to the solvable elliptic function potentials $V(x) = mj(j+1){\rm sn}^2(x,m)$ produces new exactly solvable one-dimensional periodic potentials.
\end{abstract}
\vspace{.3in}

In a recent paper, Dunne and Feinberg \cite{df} have systematically discussed various aspects of supersymmetric quantum mechanics (SUSYQM) as applied to periodic potentials. In particular, they defined and developed the concept of self-isospectral periodic potentials at length. Basically, a one dimensional potential $V_-(x)$ of period $2K$ is said to be self-isospectral if its supersymmetric partner potential $V_+(x)$ is just the original potential upto a discrete transformation - a translation by any constant amount, a reflection, or both. An example is translation by half a period, that is $V_+(x)=V_-(x-K)$. In this sense, a self-isospectral potential is somewhat trivial, since application of the SUSYQM formalism \cite{cks} to it yields nothing new. The main  example considered in ref. \cite{df} is the class of elliptic function potentials
\begin{equation}  \label{eq1}
V(x) = mj(j+1){\rm sn}^2(x,m)~~,~~j=1,2,3,\ldots
\end{equation}
Here ${\rm sn}(x,m)$ is a Jacobi elliptic function of real elliptic modulus
parameter $m$ $( 0\leq m\leq 1)$. From now on, for simplicity, the argument $m$ is suppressed. The Schr\"{o}dinger equation of the given elliptic potential is the well-known Lam\'{e}  equation \cite{ar}.
There are $j$ bound
bands (whose edges have known energies) followed by a continuum band.
In ref. \cite{df} it is claimed that the potentials given in eq. (\ref{eq1})
are self-isospectral. 
The purpose of this comment is to point out that although the $j=1$ potential is
self-isospectral, this is not the case for higher
values of $j$. Indeed, for $j \ge 2$, we claim that SUSYQM generates new exactly solvable periodic problems.

Taking the case $j = 2$, and shifting the potential by a constant so that the 
ground state has zero energy gives 
\begin{equation}
V_{-} (x) = -2-2m+2\delta+6m {\rm sn}^2(x) \, ,~\delta = \sqrt{1-m+m^2} \, .
\end{equation}
The band edge energies and Bloch wave functions $\psi_n^{(-)}(x)$ \cite{ar} are given in Table 1.
The superpotential is
\begin{equation}
W \equiv - {d\over dx} log \psi^{(-)}_{0} (x)
= {6m~{\rm sn}(x) {\rm cn}(x) {\rm dn}(x)\over
1+m+\delta-3m {\rm sn}^2 (x)} \, ,
\end{equation}
The supersymmetric partner potentials $V_{\pm} (x)$
are related to $W(x)$ via $V_{\pm} (x) = W^2 (x) \pm dW/dx.$ Hence, the potential $V_{+}$ is given by
\begin{equation} \label{e4}
V_{+}(x) = -V_-(x)+ 
{72 m^2 {\rm sn}^2(x) {\rm cn}^2(x) {\rm dn}^2(x)\over [1+m+\delta-3m {\rm sn}^2 (x)]^2}~~.
\end{equation}
Using SUSYQM and the known eigenfunctions $\psi^{(-)}_n (x)$ of $V_{-}(x)$
one can
immediately
write down the corresponding un-normalized eigenfunctions $\psi^{(+)}_n (x)$
of $V_{+} (x)$. 
\begin{equation}
\psi^{(+)}_{0} (x) = {1\over \psi^{(-)}_{0} (x)} \,~ ,~~
\psi^{(+)}_{n} (x) = ({d\over dx} +W(x)) \psi^{(-)}_{n} (x) \,~ .
\end{equation}
We have computed the band edge eigenfunctions of $V_+(x)$ and give them in Table 1. Our expression for $V_+(x)$ [eq. (\ref{e4})] does not agree with eq. (29) in ref. \cite{df}. We have checked the correctness of our results by direct substitution into the Schr\"{o}dinger equation, and by noting that 
in the limit of $m\rightarrow 1$, our $V_{+} (x)\rightarrow
4-2~{\rm sech}^2 x$, 
which indeed is the supersymmetric partner of 
$V_{-} (x,m=1)=4-6~{\rm sech}^2x~$ \cite{cks}.

Proceeding in the same way, we
have also obtained a new periodic potential
$V_{+} (x)$
corresponding to $j = 3$ case of eq. (\ref{eq1}). Here, the ground state wave function is 
$$\psi^{(-)}_{0} (x) = {\rm dn}(x) [1+2m+\delta_1-5m {\rm sn}^2(x)] \, $$
and the corresponding superpotential is
\begin{equation}
W = {m {\rm sn}(x) {\rm cn}(x)\over {\rm dn}(x)}
\ {[2m+\delta_1+11-15m {\rm sn}^2(x)]\over
[2m+\delta_1+1-5m {\rm sn}^2(x)]} \, .
\end{equation} 
The partner potentials $V_{\pm} (x)$ turn out to be
$$
V_{-}(x) = -2-5m+2\delta_1+12m {\rm sn}^2(x) ~ ,~\delta_1 \equiv \sqrt{1-m+4m^2} \, ,
$$
and
\begin{equation}
V_{+}(x) = -V_{-}(x) +{2m^2 {\rm sn}^2(x) {\rm cn}^2(x)\over
{\rm dn}^2(x)} {[2m+\delta_1+11-15m {\rm sn}^2(x)]^2\over
[2m+\delta_1+1-5m {\rm sn}^2(x)]^2} \, .
\end{equation}
Clearly, the
potential $V_{-}(x)$ is not self-isospectral.
In fact, $V_{-}(x)$ and $V_{+}(x)$ are distinctly different periodic potentials which have the same seven  
band edges corresponding to three bound bands
and a continuum band \cite{ar}. 

Although in this comment we have only focused on the $j=2,3$ cases, it is clear that SUSYQM provides a way of generating new solvable problems for all higher $j$ values. This is an exciting result given the extreme scarcity of analytically solvable periodic potentials.  Indeed, a further extension to even more general potentials involving Jacobi elliptic functions \cite{ar} yields additional quasi exactly solvable periodic potentials \cite{ks}. Partial financial support from the U.S. Department of Energy is gratefully acknowledged.


\vspace{.1in}
\centerline {Table 1: Band Edge Eigenstates of $V_{\pm}$ for $j=2~~[\delta\equiv \sqrt{1-m+m^2}~,~B\equiv
1+m+\delta]$}
\vspace{.2in}
\begin{tabular}{cccccc}
\hline
 n & $E_n$ & $\psi^{(-)}_n$ & $[B-3m~{\rm sn}^2 (x)]~\psi^{(+)}_{n}$\\
\hline
 0 & 0 & $m + 1 +\delta-3m {\rm sn}^2 (x)$  & 1  \\
 1 & $ 2\delta-1-m$ & ${\rm cn}(x) {\rm dn}(x)$
& ${\rm sn}(x)[6m-(m+1)B+m~{\rm sn}^2 (x) (2B-3-3m)]$ \\
 2 & $ 2\delta-1+2m$ & ${\rm sn}(x) {\rm dn}(x)$
& $ {\rm cn}(x) [B+m~{\rm sn}^2 (x) (3-2B)]$   & \\
 3 & $ 2\delta+2-m$ & ${\rm sn}(x) {\rm cn}(x)$
& ${\rm dn}(x) [B+{\rm sn}^2 (x) (3m-2B)]$   & \\
 4 & 4$\delta$ & $m + 1 -\delta-3m~{\rm sn}^2 (x)$ & ${\rm sn}(x){\rm cn}(x){\rm dn}(x)$   \\
\hline
\end{tabular}

\end{document}